\documentclass[letter]{aa} % for the letters 

\usepackage{graphicx}
\usepackage{txfonts}
\usepackage[]{hyperref}
\usepackage{breakurl}
\newcommand{\urlwofont}[1]{\urlstyle{same}\url{#1}}

\usepackage{lscape}

\def\FeII{Fe\,{\sc ii}} 
\def\NaI{Na\,{\sc i}} 

\begin{document}

   \title{On the triple peaks of SNHunt248 in NGC 5806}

   \subtitle{}

  \author{E.~Kankare\inst{1}
         \and
         R.~Kotak\inst{1}
         \and
         A.~Pastorello\inst{2}
         \and
         M.~Fraser\inst{3}
         \and
         S.~Mattila\inst{4}
         \and
         S.~J.~Smartt\inst{1}
         \and
         A.~Bruce\inst{5}
         \and
         K.~C.~Chambers\inst{6}
         \and
         N.~Elias-Rosa\inst{2}
         \and
         H.~Flewelling\inst{6}
         \and
         C.~Fremling\inst{7}
         \and
         J.~Harmanen\inst{8,9}
         \and
         M.~Huber\inst{6}
         \and
         A.~Jerkstrand\inst{1}
         \and
         T.~Kangas\inst{9}
         \and
         H.~Kuncarayakti\inst{10,11}
         \and
         M.~Magee\inst{1}
         \and
         E.~Magnier\inst{6}
         \and
         J.~Polshaw\inst{1}
         \and
         K.~W.~Smith\inst{1}
         \and
         J.~Sollerman\inst{7}
         \and
         L.~Tomasella\inst{2}
         }
          
   \institute{Astrophysics Research Centre, School of Mathematics and Physics, Queen's University Belfast, BT7 1NN, UK\\
              e-mail: e.kankare@qub.ac.uk
         \and
         INAF-Osservatorio Astronomico di Padova, Vicolo dell'Osservatorio 5, 35122 Padova, Italy
         \and
         Institute of Astronomy, University of Cambridge, Madingley Road, Cambridge CB3 0HA, UK
         \and
         Finnish Centre for Astronomy with ESO (FINCA), University of Turku, V{\"a}is{\"a}l{\"a}ntie 20, 21500 Piikki{\"o}, Finland
         \and
         Institute for Astronomy, University of Edinburgh, Royal Observatory, Blackford Hill, Edinburgh EH9 3HJ, UK
         \and
         Institute for Astronomy, University of Hawaii, 2680 Woodlawn Drive, Honolulu, HI 96822, USA
         \and
         Department of Astronomy, The Oskar Klein Center, Stockholm University, AlbaNova, 10691 Stockholm, Sweden
         \and
         Nordic Optical Telescope, Apartado 474, 38700 Santa Cruz de La Palma, Spain
         \and
         Tuorla Observatory, Department of Physics and Astronomy, University of Turku, V\"ais\"al\"antie 20, 21500 Piikki\"o, Finland
         \and
         Millennium Institute of Astrophysics, Casilla 36-D, Santiago, Chile
         \and
         Departamento de Astronom\'ia, Universidad de Chile, Casilla 36-D, Santiago, Chile
             }

%   \date{Received September 15, 1996; accepted March 16, 1997}

\abstract{We present our findings on a supernova (SN) impostor, SNHunt248, based on optical and near-IR data spanning $\sim$15~yrs before discovery, to $\sim$1~yr post-discovery. The light curve displays three distinct peaks, the brightest of which is at $M_{R} \sim -15.0$~mag. The post-discovery evolution is consistent with the ejecta from the outburst interacting with two distinct regions of circumstellar material. The $0.5-2.2$~$\mu$m spectral energy distribution at $-740$~d is well-matched by a single 6700~K blackbody with $\log(L/L_\sun) \sim 6.1$. This temperature and luminosity support previous suggestions of a yellow hypergiant progenitor; however, we find it to be brighter than the brightest and most massive Galactic late-F to early-G spectral type hypergiants. Overall the historical light curve displays variability of up to $\sim \pm1$~mag. At current epochs ($\sim$1~yr post-outburst), the absolute magnitude ($M_{R} \sim -9$~mag) is just below the faintest observed historical absolute magnitude $\sim$10~yrs before discovery.}


   \maketitle

\section{Introduction}

Extragalactic transient events that exhibit fainter peak absolute magnitudes than most common types of SNe have been dubbed `SN impostors' \citep{vandyk00}. Some of them have been linked to luminous blue variable (LBV) stars. By definition, these are non-terminal outbursts, thought to arise due to discrete mass-loss episodes via currently ill-understood mechanisms.

Here we report observations of a recently discovered eruptive transient in NGC~5806 \citep[$D_{L} = 22.5$~Mpc, $\mu = 31.76$~mag,][]{tully09}. Originally discovered by CRTS \citep{drake09} on 2014 May 21.18 UT, a $\gtrsim$1~mag brightening was reported shortly thereafter \citep{zheng14}, together with a spectroscopic similarity to other SN impostors. Early observations, reported by \citet{mauerhan15}, show two light curve phases (`2014a' and `2014b') with a slow rise in the 2014a phase followed by a rapid brightening to the 2014b peak. We adopt the beginning of this latter phase at JD = 2456812.35 (2014 June 3.85 UT) as our reference epoch. \citet{mauerhan15} identified a precursor at the location of SNHunt248 ($\alpha = 14^{\mathrm{h}} 59^{\mathrm{m}} 59\fs47$ and $\delta = +01\degr 54\arcmin 26$\farcs$6$, J2000) in archival HST data that led them to conclude that the star was of the cool hypergiant variety.

In what follows, we present archival data of SNHunt248 over a 15~yr period before discovery and our post-discovery photometric and spectroscopic follow-up out to about 1~yr.

\section{Observations}

\begin{figure*}
\includegraphics[width=0.706\linewidth]{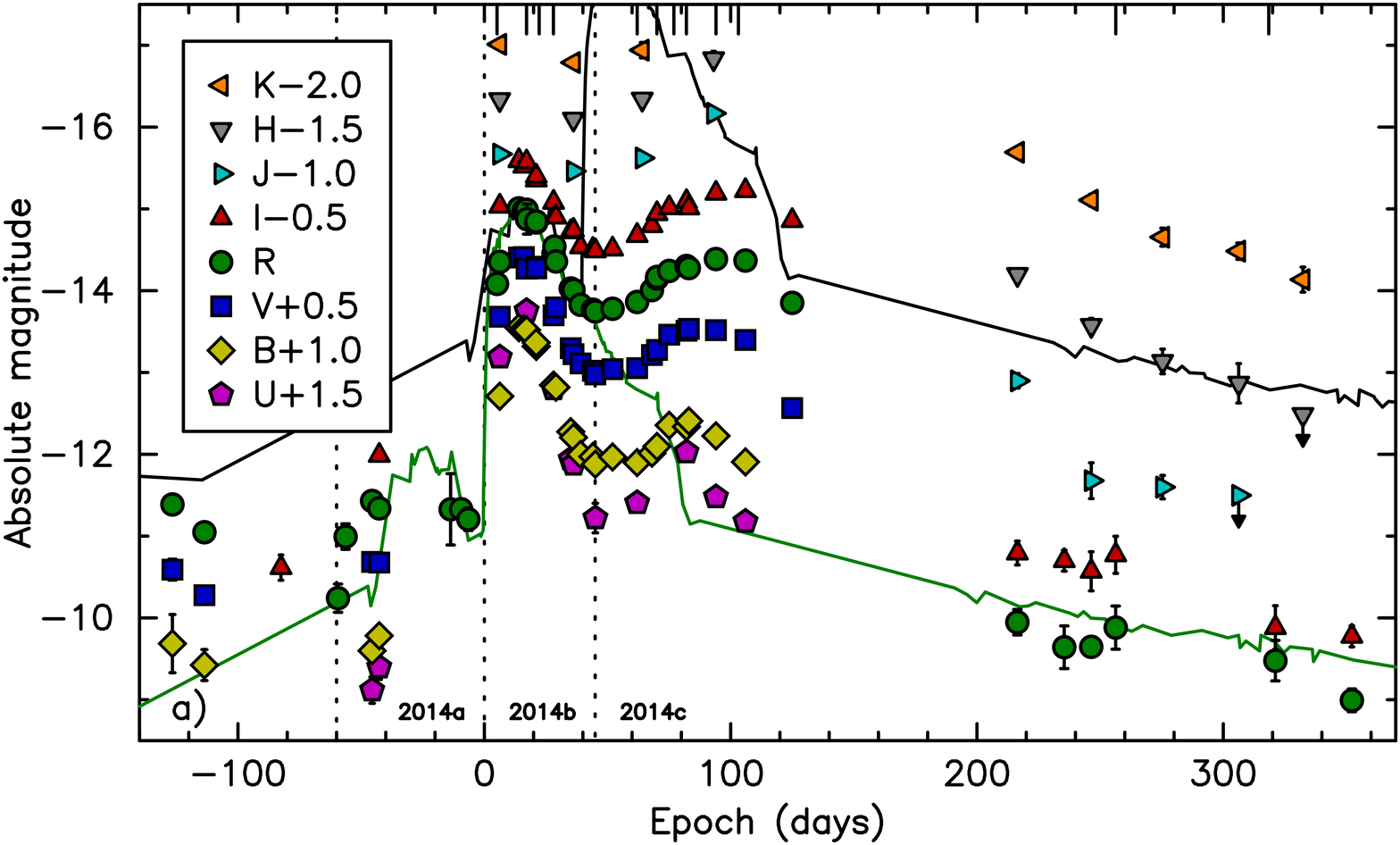}
\includegraphics[width=0.278\linewidth]{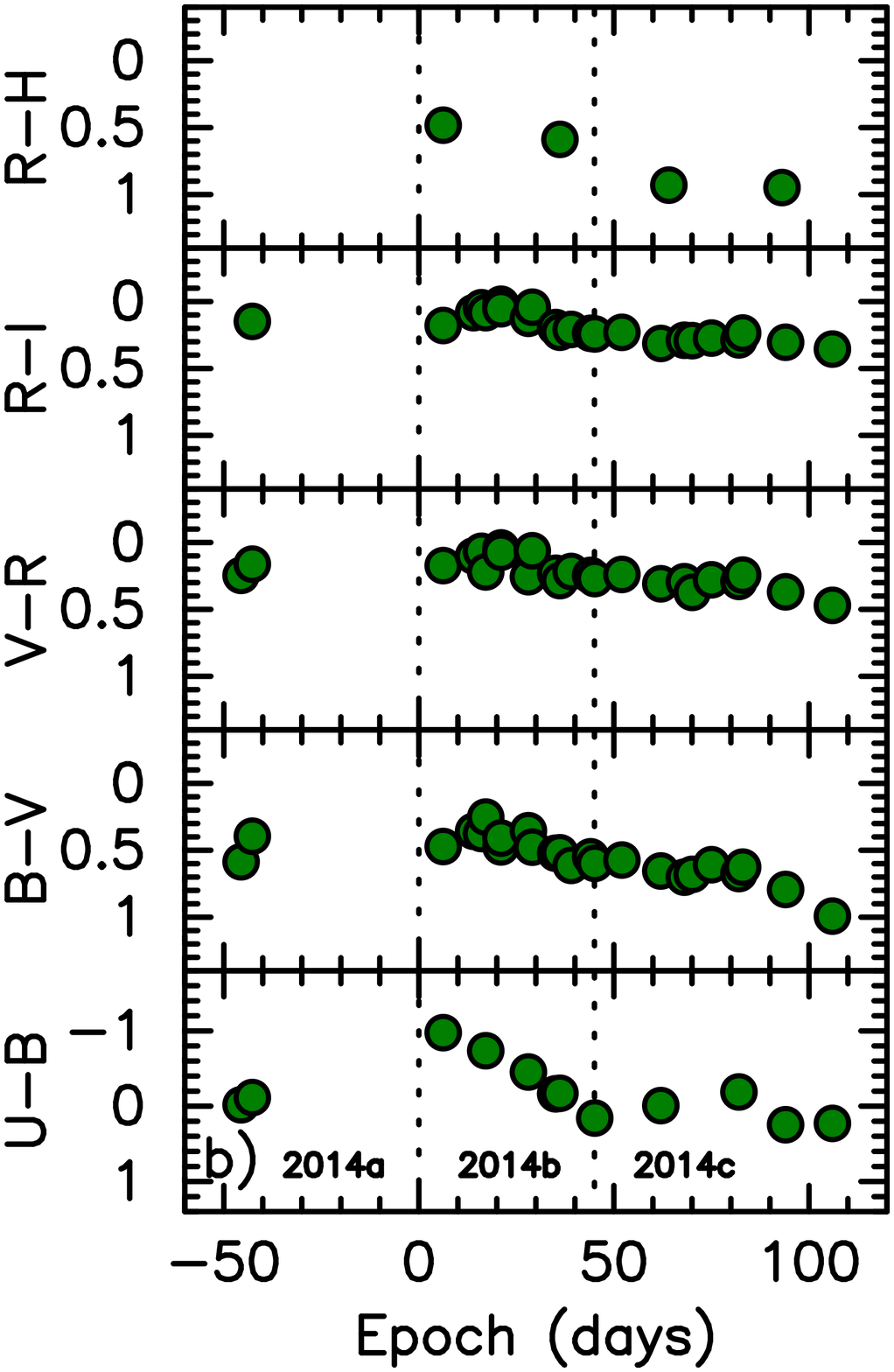}
\includegraphics[width=\linewidth]{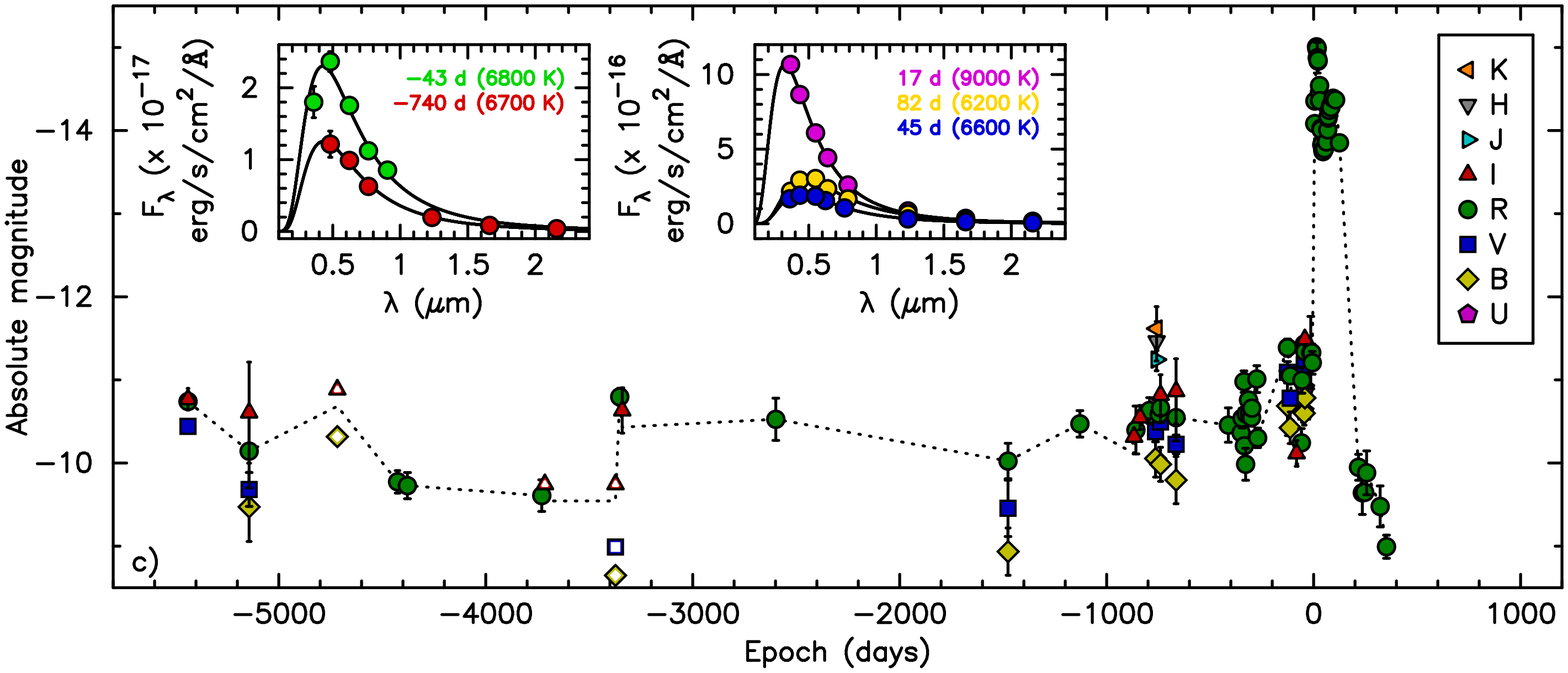}
\caption{\textbf{a)} Absolute light curves of SNHunt248. The solid vertical lines indicate the epochs of spectroscopic observations. For comparison, we show the absolute \textit{R}-band light curve of SN~2009ip \citep[][black curve: JD~$=2456155$ corresponds to day 0, no shift in magnitude; and green curve: day 0 epoch JD~$=2456195$, $+3.0$~mag shift]{pastorello13, fraser13, fraser15} emphasizing the similar evolution between some of the key phases. \textbf{b)} Colour evolution of SNHunt248. \textbf{c)} Historical light curve of SNHunt248. From day 0 onwards only the \textit{R}-band points are included for clarity. The HST archival data points from \citet{mauerhan15} are shown with open symbols. The \textit{R}-band points are connected with a dotted line to guide the eye (for epochs with an \textit{I}-band point but a missing \textit{R}-band magnitude, the latter is very roughly estimated assuming \textit{R}$-$\textit{I}~$ = 0.2$~mag, corresponding to the $-740$~d spectral energy distribution (SED)). Insets display blackbody fits to selected pre- and post-discovery epochs of observed data (interpolated post-discovery near-IR data are included for comparison).}
\label{fig:lc}
\end{figure*}

Following the reported discovery of SNHunt248, we began a follow-up campaign using several optical and near-IR facilities (see Tables~\ref{table:phot_UBVRI_hunt248} to \ref{table:spect_hunt248}). We recovered available historical light curve information on the precursor of SNHunt248 from publicly-available archival images of the field. Standard IRAF-based procedures were followed for bias subtraction and flat-field correction of the optical follow-up images and carried out using primarily the QUBA pipeline \citep{valenti11}. Processing of the near-IR follow-up frames included additional steps involving sky subtraction, image alignment and stacking using a slightly modified version of the IRAF based {\sc notcam} package. PSF fitting photometry was carried out within the QUBA pipeline, which also interpolates the sky background at the transient location based on the surrounding environment. The optical photometry was calibrated based on the SDSS DR9 \textit{ugriz} magnitudes of 15 field stars. For the Johnson-Cousins filters, we used the transformations of \citet{jester05} to convert the sequence star magnitudes to the \textit{UBVRI} system. The near-IR images were calibrated using 2MASS \textit{JHK} magnitudes of 10 sequence stars. Average instrument-specific colour terms were derived from standard star observations and applied only for optical follow up, otherwise zero colour terms were assumed. The resulting photometry is presented in Tables~\ref{table:phot_UBVRI_hunt248} to \ref{table:phot_JHK_hunt248} and in Fig.~\ref{fig:lc}.

\begin{figure*}
\includegraphics[width=\linewidth]{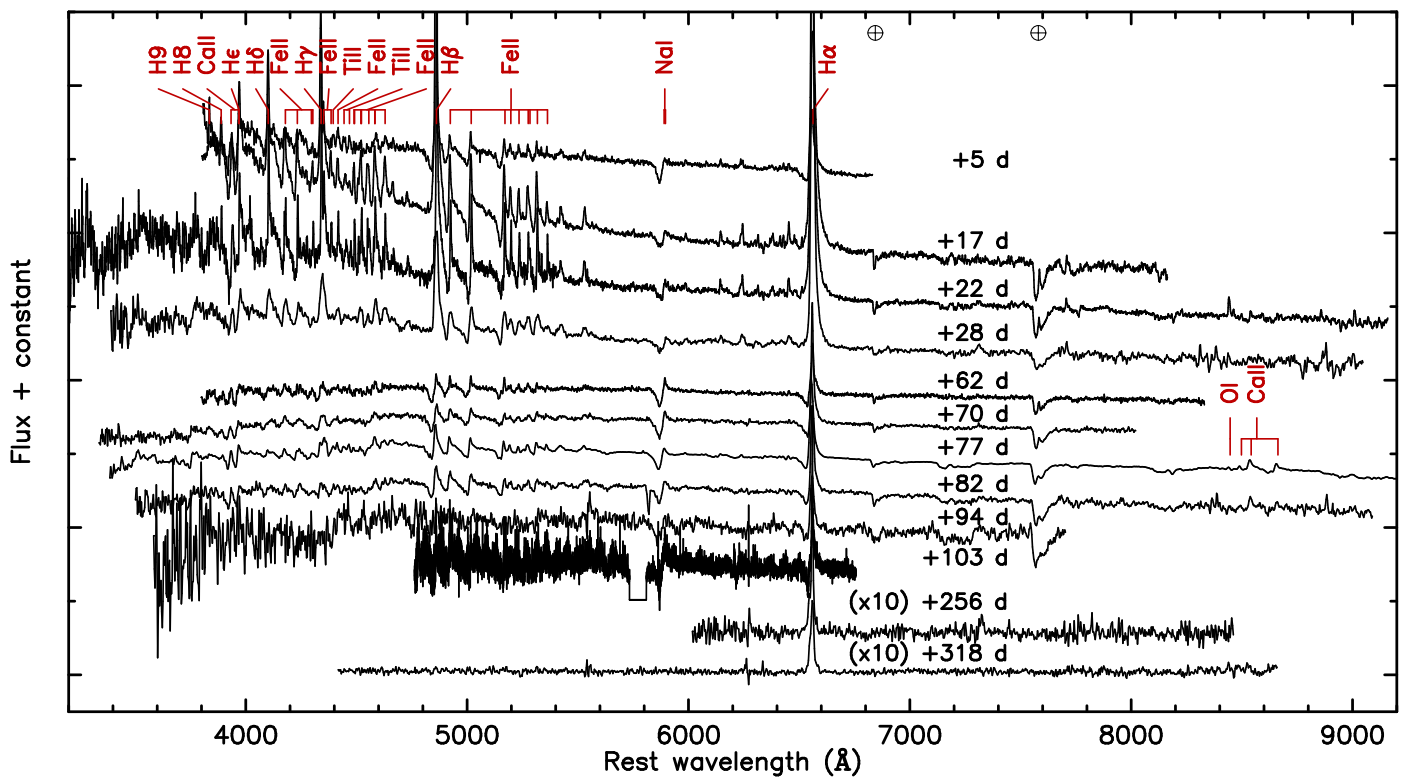}
\caption{Spectral time series of SNHunt248. The spectra have been dereddened, redshift corrected to the rest frame and vertically shifted for clarity. The strongest telluric features are indicated by $\oplus$ symbols. The identification of the most prominent lines is based on that of SN~2009kn \citep[see][and references therein]{kankare12}.}
\label{fig:spect}
\end{figure*}

Spectroscopic observations from the NOT, WHT, and TNG were also reduced using the QUBA pipeline and included the standard steps of bias subtraction, flat field correction, spectrum extraction, wavelength and relative flux calibration. Wavelength calibration was derived based on arc lamp exposures. Relative flux calibration was carried out using a sensitivity curve derived with observations of spectrophotometric standard stars obtained on the same night and setup as SNHunt248. The GTC+OSIRIS and VLT+FORS2 spectra were reduced in a similar way using basic IRAF tasks. The VLT+UVES spectrum was reduced using the Reflex-based pipeline workflow. Furthermore, all the spectra were absolute flux calibrated based on the optical photometry of SNHunt248. The spectroscopic observations are summarized in Table~\ref{table:spect_hunt248} and the spectral sequence of SNHunt248 is shown in Fig.~\ref{fig:spect}.

\section{Analysis}

We adopt $A_{V}^{\mathrm{Gal}} = 0.140$~mag for Galactic extinction towards SNHunt248 \citep{schlafly11}. In our highest resolution spectrum (+103~d, R = 26000), we do not detect the narrow interstellar medium (ISM) absorption components of the \NaI~D at the expected redshift of NGC~5806. SNHunt248 lies at a relatively large projected offset ($\sim$6.4~kpc) from the host nucleus in an uncrowded region. This is suggestive of small host galaxy reddening and we adopt $A_{V}^{\mathrm{host}} = 0$~mag. 

The most striking feature in the post-discovery evolution of SNHunt248 is its triple-peaked light curve. Our data shows a clear decline in the \textit{R}-band light curve between $-127$ and $-60$~d which marks the beginning of the rise to the phase 2014a, with the first peak at $M_{R} \approx -11.4$~mag. In the 2014b phase the light curve rises $>$3~mag to the second peak (and maximum) in $\sim$15 days in all the optical bands, followed by a decline of $\sim$1~mag for the next $\sim$30 days. This is followed by a phase of re-brightening. The rise to the third maximum is less rapid and takes $\sim$50 d, with the blue bands evolving faster than the red ones. Extending the naming convention of \citet{mauerhan15}, we dub this third light curve phase as `2014c'. The absolute magnitude of this peak ($M_{R}\approx-14.4$ mag) does not reach the brightness of the 2014b maximum ($M_{R}\approx-15.0$ mag) which is somewhat brighter than the light curve maximum of most events classified as SN impostors \citep{smith11}, but fainter than normal SNe. SNHunt248 colour is bluest at the peak of the phase 2014b peak, followed by a fairly linear evolution towards a redder colour (see Fig.~\ref{fig:lc}). By $\sim$~+250~d SNHunt248 has already reached the historical light curve minima observed $\sim$3000 to 7000~d before the outburst. Overall the historical light curve of SNHunt248 has shown variability of up to $\pm1$~mag over the last 1.5 decades.

\begin{figure}
\includegraphics[width=\linewidth]{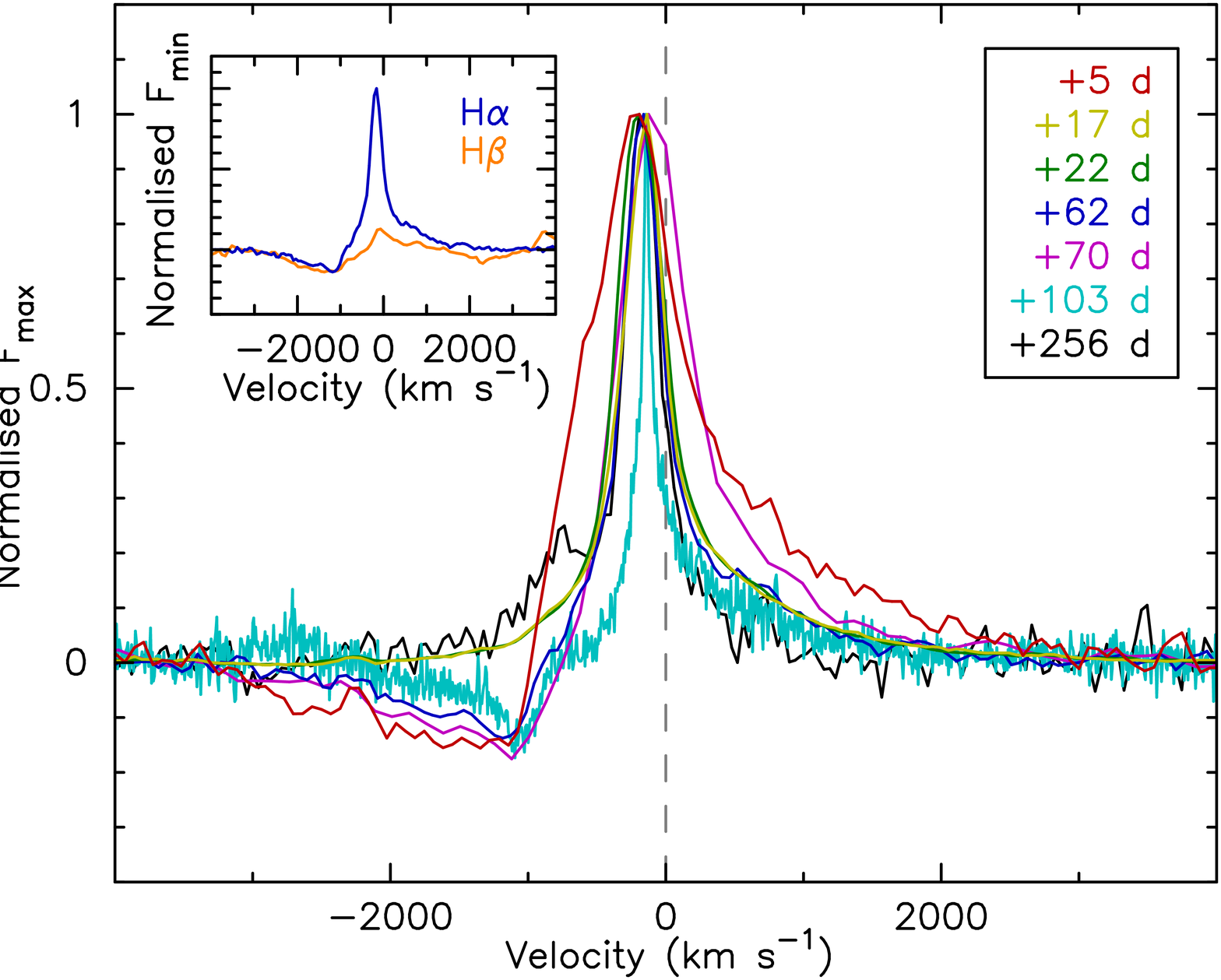}
\caption{Selection of continuum-subtracted and peak-normalised H$\alpha$ profiles of SNHunt248. For clarity, only spectra with R~$> 500$ are shown. The inset shows the minimum-normalised +62~d H$\alpha$ and H$\beta$ profiles with identical absorption components. The narrow emission component is resolved only in the +103~d VLT+UVES spectrum.}
\label{fig:Halpha}
\end{figure}

Our earliest spectrum (+5~d, phase 2014b) shows P Cygni line profiles of Balmer lines and a forest of \FeII\ and other metal lines with a strong absorption. The Balmer lines are the most prominent features in the spectra of SNHunt248. The H$\alpha$ absorption minimum indicates a velocity of $\sim$1100~km~s$^{-1}$ while the blue edge of the asymmetric absorption component extends to $\sim$3000~km~s$^{-1}$. No strong evolution of the metal lines is seen whereas the most notable evolution in the line profiles is the disappearance of the absorption features in the Balmer lines by the time the light curve reaches the phase 2014b maximum. The absorption feature in the Balmer lines remains absent throughout the light curve decline of this phase. However, as the light curve evolves into the rise towards phase 2014c, the absorption features reappear in the Balmer lines (Fig.~\ref{fig:Halpha}). It is noteworthy that the absorption component at +62 and +70~d, i.e. in phase 2014c, displays an essentially identical absorption component profile compared to the early +5~d spectrum, indicating material moving at the same velocity as in phase 2014b. Unlike the Balmer lines, the metal lines show a P Cygni profile throughout our observed time series from the early 2014b phase to the 2014c peak, but become less prominent and in the late-time spectra the low signal-to-noise ratio prevents identification of all but the strongest lines. 

The field of SNHunt248 was observed $\sim$2~yrs before the 2014 discovery with WHT+LIRIS in \textit{JHK} and GTC+OSIRIS in \textit{gri} within 19 days. The SED is well described by a single blackbody yielding $T_{\mathrm{BB}} \approx 6700$~K, $R_{\mathrm{BB}} \approx  5.9 \times 10^{13}$~cm ($\sim 850$~$R_{\sun}$), and $L_{\mathrm{BB}} \approx  5.0 \times 10^{39}$~erg~s$^{-1}$ ($\sim 1.3 \times 10^{6}$~$L_{\sun}$), see Fig.~\ref{fig:lc}. This would suggest the SNHunt248 progenitor to be a yellow hypergiant (YHG) close to the `Yellow Void' \citep{dejager98}. YHGs are unstable and eruptive stars, and some of them have dusty circumstellar envelopes. The SNHunt248 SED shows no sign of a near-IR excess at day $-740$ out to 2.2~$\mu$m. Furthermore, the variable luminosity of SNHunt248 at $-740$~d is above, albeit only briefly, the Humphreys-Davidson limit ($\log L \approx 5.8$~$L_{\sun}$ at the corresponding $T$) where massive LBVs in outburst are often found. In the coolest eruptive states, LBVs can resemble type F hypergiants with temperatures of $\sim$7000$-$8000~K \citep{humphreys94}. The SNHunt248 precursor is close, but below this temperature lower limit. For comparison, \citet{mauerhan15} estimated $T \approx 6500$~K, $R \approx 500$~$R_{\odot}$, and $L \approx 4 \times 10^{5}$~$L_{\odot}$ for the SNHunt248 precursor on $-3373$~d. This is roughly the observed range of the pre-discovery variability, prominent mainly in the radius and luminosity. It is possible that the observed variability in the historical light curve is arising some or \textit{all} the times from discrete mass-loss episodes, possibly related to the formation of a pseudo-photosphere.

Based on our early photometry, the initial outburst of SNHunt248, and the rise to phase 2014a, commenced at $\sim -60$~d, followed by a brightening of the light curves to the double-peaked maximum powered by strong circumstellar medium (CSM) interaction. This is consistent with the early ($-9$~d) spectrum, reported by \citet{mauerhan15}, displaying similar P Cygni velocities \textit{before} phase 2014b as seen during phase 2014b and 2014c, since the P Cygni absorption component likely arise from the outburst. It is therefore also unlikely that the 2014c phase arose from a new major outburst of the progenitor since no evolution is observed in the velocities of the H$\alpha$ absorption component. A more likely explanation is that the outburst ejecta from the 2014a event collided into two shell-like regions of dense CSM, powering phases 2014b and 2014c. A possible Galactic YHG analogue to this scenario would be IRAS~17163-3907 which has been shown to have two shell-like structures at $3.6 \times 10^{16}$ and $9 \times 10^{16}$~cm \citep{lagadec11}. Differences in the light curve and spectral evolution of phases 2014b and 2014c likely arise because the CSM shells are not identical in terms of their masses, densities, and velocities. The narrow emission component of H$\alpha$ is resolved in our +103~d (phase 2014c) spectrum; a Lorentzian fit yields a FWHM of $\sim$90~km~s$^{-1}$, which is consistent with stellar winds of many LBVs and also some YHGs \citep{lagadec11}. 

We assume an initial radius of $6 \times 10^{13}$~cm from our $-740$~d data, the initial eruption already taking place at $-60$~d with a bulk expansion velocity of $\sim$1100~km~s$^{-1}$. The rise to the next maximum (phase 2014b) results in $\sim6.4$ to $\sim7.7 \times 10^{14}$~cm and phase 2014c maximum from $\sim1.1$ to $\sim1.5 \times 10^{15}$~cm. During the 2014b phase we do not resolve the narrow emission component; however, the resolution of our WHT spectrum allows us to set an upper limit of $<300$~km~s$^{-1}$ for the CSM wind, suggesting a period of mass-loss $\gtrsim$0.7~yr before the outburst. For phase 2014c, the 90~km~s$^{-1}$ CSM wind suggest the earlier mass-loss episode to have taken place from $\sim$3.5 to $\sim$5~yrs before the events of 2014.

\section{Conclusions}

We find the unique 2014 triple-peaked brightening of SNHunt248 to be arising from an outburst of a progenitor with an appearance of a very luminous YHG with no clear near-IR excess. With an optical peak magnitude of roughly $-15$~mag, the SNHunt248 event is among the most luminous SN impostors. The historical light curve of SNHunt248 presented here offers a rare pre-discovery view to the evolution of an SN impostor precursor and shows variability episodes. The photometric and spectroscopic evolution of SNHunt248 is consistent with the outburst ejecta interacting with two distinct regions of CSM giving arise to a luminous display, not observed in Galactic YHGs. 

\begin{acknowledgements}

We thank the anonymous referee for useful comments and Justyn Maund for helpful discussions.

We are grateful to Kathleen Eckert and co-authors for providing early access to DECam \textit{ugriz} data for NGC~5806 (NOAO 2014A-0256). We thank Stefano Benetti and Leonardo Tartaglia for their help with some of the observations.

RK and SJS acknowledge support from STFC (ST/L000709/1). SJS acknowledges funding from the European Research Council under the European Union’s Seventh Framework Programme (FP7/2007-2013)/ERC grant agreement no. [291222]. This work was partly supported by the European Union FP7 programme through ERC grant number 320360. AP and LT are partially supported by the PRIN-INAF 2012 with the project ``Transient Universe: from ESO Large to PESSTO''. NER acknowledges the support from the European Union Seventh Framework Programme (FP7/2007-2013) under grant agreement no. 267251 ``Astronomy Fellowships in Italy'' (AstroFIt). Support for HK is provided by the Ministry of Economy, Development, and Tourism's Millennium Science Initiative through grant IC120009, awarded to The Millennium Institute of Astrophysics, MAS. HK acknowledges support by CONICYT through FONDECYT grant 3140563.

This work is based on observations made at the following facilities: CFHT, CRTS, CTIO, ESO (programme IDs 091.D-0516, 093.D-0798, 095.D-0793, 177.A-3016, and 184.D-1151), FTN, Gemini, GTC, INAF, ING, KPNO, LT, NOT, PS1, SDSS, and TNG.  

The Pan-STARRS1 Surveys (PS1) have been made possible through contributions of the Institute for Astronomy, the University of Hawaii, the Pan-STARRS Project Office, the Max-Planck Society and its participating institutes, the Max Planck Institute for Astronomy, Heidelberg and the Max Planck Institute for Extraterrestrial Physics, Garching, The Johns Hopkins University, Durham University, the University of Edinburgh, Queen's University Belfast, the Harvard-Smithsonian Center for Astrophysics, the Las Cumbres Observatory Global Telescope Network Incorporated, the National Central University of Taiwan, the Space Telescope Science Institute, the National Aeronautics and Space Administration under Grant No. NNX08AR22G issued through the Planetary Science Division of the NASA Science Mission Directorate, the National Science Foundation under Grant No. AST-1238877, the University of Maryland, and Eotvos Lorand University (ELTE).

Operation of the Pan-STARRS1 telescope is supported by the National Aeronautics and Space Administration under Grant No. NNX12AR65G and Grant No. NNX14AM74G issued through the NEO Observation Program. 

\end{acknowledgements}

\newpage

\Online

\onllongtab{
\begin{longtable}{ccccccccc}
\caption{Optical \textit{UBVRI} photometry for SNHunt248 with the errors given in brackets.} \\
\hline
\hline
JD & $t$ & $m_{U}$ & $m_{B}$ & $m_{V}$ & $m_{R}$ & $m_{I}$ & Telescope+instrument & Seeing\\
(2400000+) & (d) & (mag) & (mag) & (mag) & (mag) & (mag) & & (\arcsec)\\ 
\hline
51373.79 & -5439 & $\ldots$ & $\ldots$ & 21.458(0.053) & 21.138(0.058) & 21.076(0.130) & CFHT+CFH12K & 0.9\\
51668.89 & -5143 & $\ldots$ & 22.475(0.416) & 22.219(0.204) & 21.736(0.442) & 21.237(0.608) & SDSS$^{\mathrm{a}}$ & 1.0\\
52386.06 & -4426 & $\ldots$ & $\ldots$ & $\ldots$ & 22.104(0.135) & $\ldots$ & CFHT+CFH12K & 1.2\\
52433.98 & -4378 & $\ldots$ & $\ldots$ & $\ldots$ & 22.150(0.157) & $\ldots$ & CFHT+CFH12K & 1.2\\
53082.70 & -3730 & $\ldots$ & $\ldots$ & $\ldots$ & 22.270(0.190) & $\ldots$ & INT+WFC$^{\mathrm{b}}$ & 1.2\\
53459.66 & -3353 & $\ldots$ & $\ldots$ & $\ldots$ & 21.082(0.103) & $\ldots$ & INT+WFC & 1.2\\
53471.81 & -3341 & $\ldots$ & $\ldots$ & $\ldots$ & $\ldots$ & 21.212(0.272) & CTIO4m+Mosaic-2 & 1.6\\
54213.81 & -2599 & $\ldots$ & $\ldots$ & $\ldots$ & 21.352(0.254) & $\ldots$ & KPNO4m+Mosaic-1 & 1.9\\
55335.77 & -1477 & $\ldots$ & 23.013(0.285) & 22.445(0.339) & 21.854(0.214) & $\ldots$ & KPNO4m+Mosaic-1 & 2.8\\
55682.95 & -1129 & $\ldots$ & $\ldots$ & $\ldots$ & 21.407(0.159) & $\ldots$ & PS1+GPC1$^{\mathrm{c}}$ & 1.3\\
55944.65 & -868 & $\ldots$ & $\ldots$ & $\ldots$ & $\ldots$ & 21.528(0.077) & Gemini-N+GMOS$^{\mathrm{d}}$ & 0.7\\
55953.70 & -859 & $\ldots$ & $\ldots$ & $\ldots$ & 21.481(0.288) & $\ldots$ & LT+RATCam$^{\mathrm{b}}$ & 1.2\\
55973.86 & -838 & $\ldots$ & $\ldots$ & $\ldots$ & $\ldots$ & 21.296(0.144) & NTT+EFOSC2 & 0.9\\
56019.92 & -792 & $\ldots$ & $\ldots$ & $\ldots$ & 21.246(0.154) & $\ldots$ & FTN+FS02 & 2.1\\
56048.57 & -764 & $\ldots$ & 21.893(0.223) & 21.523(0.121) & $\ldots$ & $\ldots$ & NOT+ALFOSC$^{\mathrm{a}}$ & 0.9\\
56065.89 & -746 & $\ldots$ & $\ldots$ & $\ldots$ & 21.291(0.083) & $\ldots$ & PS1+GPC1$^{\mathrm{c}}$ & 0.9\\
56072.64 & -740 & $\ldots$ & 21.963(0.205) & 21.405(0.089) & 21.215(0.179) & 21.025(0.244) & GTC+OSIRIS$^{\mathrm{a}}$ & 1.0\\
56147.54 & -665 & $\ldots$ & 22.153(0.283) & 21.675(0.112) & 21.332(0.280) & 20.980(0.390) & NTT+EFOSC2$^{\mathrm{a}}$ & 1.3\\
56400.03 & -412 & $\ldots$ & $\ldots$ & $\ldots$ & 21.421(0.207) & $\ldots$ & PS1+GPC1$^{\mathrm{c}}$ & 1.5\\
56461.44 & -351 & $\ldots$ & $\ldots$ & $\ldots$ & 21.514(0.176) & $\ldots$ & TNG+DOLORES & 1.1\\
56466.50 & -346 & $\ldots$ & $\ldots$ & $\ldots$ & 21.339(0.117) & $\ldots$ & VLT+FORS2$^{\mathrm{e}}$ & 0.8\\
56475.89 & -336 & $\ldots$ & $\ldots$ & $\ldots$ & 20.899(0.130) & $\ldots$ & FTN+FS02 & 2.4\\
56477.49 & -335 & $\ldots$ & $\ldots$ & $\ldots$ & 21.671(0.087) & $\ldots$ & VLT+FORS2$^{\mathrm{e}}$ & 0.5\\
56483.57 & -329 & $\ldots$ & $\ldots$ & $\ldots$ & 21.893(0.192) & $\ldots$ & VLT+FORS2$^{\mathrm{e}}$ & 0.9\\
56485.56 & -327 & $\ldots$ & $\ldots$ & $\ldots$ & 21.287(0.063) & $\ldots$ & VLT+FORS2$^{\mathrm{e}}$ & 0.5\\
56497.47 & -315 & $\ldots$ & $\ldots$ & $\ldots$ & 21.120(0.064) & $\ldots$ & VLT+FORS2$^{\mathrm{e}}$ & 0.8\\
56501.50 & -311 & $\ldots$ & $\ldots$ & $\ldots$ & 21.287(0.135) & $\ldots$ & VLT+FORS2$^{\mathrm{e}}$ & 1.0\\
56512.52 & -300 & $\ldots$ & $\ldots$ & $\ldots$ & 21.331(0.104) & $\ldots$ & VLT+FORS2$^{\mathrm{e}}$ & 0.9\\
56513.49 & -299 & $\ldots$ & $\ldots$ & $\ldots$ & 21.217(0.081) & $\ldots$ & VLT+FORS2$^{\mathrm{e}}$ & 0.7\\
56537.48 & -275 & $\ldots$ & $\ldots$ & $\ldots$ & 20.868(0.162) & $\ldots$ & VLT+FORS2$^{\mathrm{e}}$ & 1.4\\
56541.47 & -271 & $\ldots$ & $\ldots$ & $\ldots$ & 21.576(0.117) & $\ldots$ & VLT+FORS2$^{\mathrm{e}}$ & 0.7\\
56685.73 & -127 & $\ldots$ & 21.261(0.356) & 20.809(0.128) & 20.490(0.104) & $\ldots$ & NOT+ALFOSC$^{\mathrm{a,b}}$ & 1.4\\
56698.72 & -114 & $\ldots$ & 21.523(0.190) & 21.121(0.076) & 20.830(0.083) & $\ldots$ & NOT+ALFOSC$^{\mathrm{a,b}}$ & 1.0\\
56729.81 & -83 & $\ldots$ & $\ldots$ & $\ldots$ & $\ldots$ & 21.728(0.155) & VST+OmegaCAM$^{\mathrm{d}}$ & 0.6\\
56752.95 & -59 & $\ldots$ & $\ldots$ & $\ldots$ & 21.636(0.173) & $\ldots$ & PS1+GPC1$^{\mathrm{c}}$ & 1.2\\
56756.01 & -56 & $\ldots$ & $\ldots$ & $\ldots$ & 20.883(0.156) & $\ldots$ & PS1+GPC1$^{\mathrm{c}}$ & 1.5\\
56766.79 & -46 & 21.362(0.161) & 21.348(0.091) & 20.715(0.045) & 20.447(0.047) & $\ldots$ & SOAR+Goodman & 1.2\\
56769.68 & -43 & 21.081(0.150) & 21.164(0.086) & 20.722(0.032) & 20.538(0.053) & 20.356(0.069) & CTIO4m+DECam$^{\mathrm{a}}$ & 1.0\\
56798.68 & -14 & $\ldots$ & $\ldots$ & $\ldots$ & 20.550(0.436) & $\ldots$ & MLS$^{\mathrm{f}}$ & 2.5\\
56802.91 & -9 & $\ldots$ & $\ldots$ & $\ldots$ & 20.549(0.104) & $\ldots$ & PS1+GPC1$^{\mathrm{c}}$ & 1.4\\
56805.85 & -7 & $\ldots$ & $\ldots$ & $\ldots$ & 20.672(0.137) & $\ldots$ & PS1+GPC1$^{\mathrm{c}}$ & 1.5\\
56817.47 & 5 & $\ldots$ & $\ldots$ & $\ldots$ & 17.794(0.021) & $\ldots$ & NOT+ALFOSC & 0.6\\
56818.60 & 6 & 17.292(0.107) & 18.237(0.071) & 17.718(0.038) & 17.523(0.032) & 17.310(0.025) & NOT+StanCam & 0.9\\
56826.39 & 14 & $\ldots$ & 17.399(0.027) & 16.994(0.011) & 16.871(0.025) & 16.754(0.034) & LT+IO:O$^{\mathrm{a}}$ & 0.9\\
56828.40 & 16 & $\ldots$ & 17.410(0.023) & 16.988(0.010) & 16.900(0.018) & 16.820(0.023) & LT+IO:O$^{\mathrm{a}}$ & 0.9\\
56829.49 & 17 & 16.724(0.030) & 17.428(0.012) & 17.129(0.012) & 16.887(0.011) & 16.770(0.020) & NOT+ALFOSC & 1.2\\
56829.68 & 17 & $\ldots$ & $\ldots$ & $\ldots$ & 17.002(0.184) & $\ldots$ & MLS$^{\mathrm{f}}$ & 2.3\\
56833.40 & 21 & $\ldots$ & 17.628(0.037) & 17.106(0.013) & 17.043(0.019) & 16.989(0.024) & LT+IO:O$^{\mathrm{a}}$ & 1.4\\
56833.41 & 21 & $\ldots$ & 17.582(0.016) & 17.126(0.007) & 17.031(0.014) & 16.943(0.019) & WHT+ACAM$^{\mathrm{a}}$ & 1.5\\
56840.44 & 28 & 17.680(0.032) & 18.101(0.027) & 17.700(0.017) & 17.422(0.014) & 17.263(0.020) & NOT+ALFOSC & 1.2\\
56840.80 & 28 & $\ldots$ & $\ldots$ & $\ldots$ & 17.337(0.016) & $\ldots$ & PS1+GPC1$^{\mathrm{c}}$ & 1.9\\
56841.41 & 29 & $\ldots$ & 18.127(0.030) & 17.603(0.013) & 17.519(0.025) & 17.443(0.033) & LT+IO:O$^{\mathrm{a}}$ & 1.7\\
56847.42 & 35 & 18.539(0.070) & 18.671(0.062) & 18.102(0.022) & 17.850(0.024) & 17.622(0.032) & NOT+ALFOSC & 1.6\\
56848.54 & 36 & 18.602(0.081) & 18.741(0.039) & 18.175(0.021) & 17.868(0.020) & 17.605(0.051) & NOT+StanCam & 1.3\\
56851.41 & 39 & $\ldots$ & 18.948(0.045) & 18.288(0.020) & 18.049(0.039) & 17.807(0.053) & LT+IO:O$^{\mathrm{a}}$ & 0.6\\
56856.39 & 44 & $\ldots$ & 18.974(0.027) & 18.378(0.011) & 18.110(0.027) & 17.839(0.038) & LT+IO:O$^{\mathrm{a}}$ & 0.9\\
56857.42 & 45 & 19.258(0.181) & 19.068(0.040) & 18.423(0.023) & 18.131(0.037) & 17.855(0.046) & AS1.82m+AFOSC$^{\mathrm{a}}$ & 1.8\\
56864.39 & 52 & $\ldots$ & 18.979(0.043) & 18.359(0.017) & 18.098(0.032) & 17.834(0.043) & LT+IO:O$^{\mathrm{a}}$ & 0.7\\
56874.37 & 62 & 19.072(0.049) & 19.048(0.026) & 18.346(0.015) & 18.015(0.014) & 17.668(0.020) & NOT+ALFOSC & 1.1\\
56880.41 & 68 & $\ldots$ & 18.933(0.082) & 18.185(0.031) & 17.870(0.047) & 17.547(0.060) & LT+IO:O$^{\mathrm{a}}$ & 1.2\\
56882.43 & 70 & $\ldots$ & 18.850(0.031) & 18.123(0.013) & 17.728(0.013) & 17.402(0.015) & TNG+DOLORES & 1.5\\
56882.50 & 70 & $\ldots$ & $\ldots$ & $\ldots$ & 17.709(0.012) & $\ldots$ & CTIO4m+DECam$^{\mathrm{b}}$ & 1.0\\
56887.38 & 75 & $\ldots$ & 18.592(0.030) & 17.937(0.013) & 17.636(0.023) & 17.329(0.031) & LT+IO:O$^{\mathrm{a}}$ & 1.0\\
56894.41 & 82 & 18.452(0.045) & 18.609(0.020) & 17.888(0.014) & 17.575(0.015) & 17.255(0.023) & NOT+ALFOSC & 1.0\\
56895.37 & 83 & $\ldots$ & 18.539(0.030) & 17.866(0.014) & 17.600(0.026) & 17.331(0.036) & LT+IO:O$^{\mathrm{a}}$ & 0.9\\
56906.36 & 94 & 19.000(0.065) & 18.721(0.020) & 17.881(0.016) & 17.490(0.012) & 17.152(0.014) & NOT+ALFOSC & 1.0\\
56918.35 & 106 & 19.301(0.063) & 19.040(0.025) & 18.001(0.014) & 17.509(0.015) & 17.120(0.019) & NOT+ALFOSC & 1.2\\
56937.32 & 125 & $\ldots$ & $\ldots$ & 18.834(0.077) & 18.026(0.033) & 17.486(0.029) & NOT+StanCam & 1.2\\
57028.75 & 216 & $\ldots$ & $\ldots$ & $\ldots$ & 21.930(0.156) & 21.550(0.146) & NOT+StanCam & 0.6\\
57047.78 & 235 & $\ldots$ & $\ldots$ & $\ldots$ & 22.236(0.261) & 21.643(0.128) & NOT+ALFOSC & 0.9\\
57058.71 & 246 & $\ldots$ & $\ldots$ & $\ldots$ & 22.231(0.085) & 21.773(0.239) & NOT+StanCam & 0.6\\
57068.65 & 256 & $\ldots$ & $\ldots$ & $\ldots$ & 21.996(0.264) & 21.572(0.227) & WHT+ACAM & 1.3\\
57133.49 & 321 & $\ldots$ & $\ldots$ & $\ldots$ & 22.400(0.247) & 22.461(0.267) & WHT+ACAM & 1.1\\
57164.53 & 352 & $\ldots$ & $\ldots$ & $\ldots$ & 22.885(0.139) & 22.568(0.129) & VLT+FORS2 & 0.9\\
\hline
\label{table:phot_UBVRI_hunt248}
\end{longtable}
\tablefoot{Seeing is derived from the image most similar to \textit{R}-band, and corresponds to the average FWHM of a selection of sequence stars. CFHT = 3.58-m Canada-France-Hawaii Telescope (Mauna Kea, Hawaii); SDSS = 2.5-m Sloan Digital Sky Survey telescope (Apache Point Observatory, New Mexico); INT = 2.5-m Isaac Newton Telescope (La Palma, Spain); CTIO4m = 4-m Blanco Telescope (Cerro Tololo, Chile); KPNO4m = 4-m Mayall Telescope (Kitt Peak, Arizona); PS1 = 1.8-m Panoramic Survey Telescope \& Rapid Response System 1 (Haleakala, Hawaii); Gemini-N = 8.1-m Gemini North Telescope (Mauna Kea, Hawaii); LT  = 2.0-m Liverpool Telescope (La Palma, Spain); NTT = 3.58-m ESO New Technology Telescope (La Silla, Chile); FTN = 2.0-m Faulkes Telescope North (Haleakala, Hawaii); NOT = 2.56-m Nordic Optical Telescope (La Palma, Spain); GTC = 10.4-m Gran Telescopio Canarias (La Palma, Spain); TNG = 3.58-m Telescopio Nazionale Galileo (La Palma, Spain); VLT = 8.2-m Very Large Telescope (Cerro Paranal, Chile); VST = 2.6-m VLT Survey Telescope (Cerro Paranal, Chile); SOAR = 4.1-m Southern Astrophysical Research Telescope (Cerro Pach\'on, Chile); MLS = 1.5-m Mt. Lemmon Survey Telescope (Steward Observatory, Arizona); AS1.82-m = 1.82-m Copernico Telescope (Asiago, Italy); WHT = 4.2-m William Herschel Telescope (La Palma, Spain). $^{\mathrm{a}}$ Table~\ref{table:phot_gri_hunt248} \textit{ugri} magnitudes converted to the \textit{UBVRI} system using the transformations of \citet{jester05}. $^{\mathrm{b}}$ \textit{r}-band image calibrated directly into \textit{R}. $^{\mathrm{c}}$ PS1 \textit{w}$_{\mathrm{P1}}$ image calibrated directly into \textit{R}. $^{\mathrm{d}}$ \textit{i}-band image calibrated directly into \textit{I}. $^{\mathrm{e}}$ VLT blocking filter GG435 image calibrated directly into \textit{R}. $^{\mathrm{f}}$ Unfiltered image calibrated directly into \textit{R}.}
}

\onltab{
\begin{table*}[h]
\caption{Original \textit{ugriz} photometry for SNHunt248 with the errors given in brackets.}
\centering
\begin{tabular}{ccccccccc}
\hline
\hline
JD & $t$ & $m_{u}$ & $m_{g}$ & $m_{r}$ & $m_{i}$ & $m_{z}$ & Telescope+instrument & Seeing\\
(2400000+) & (d) & (mag) & (mag) & (mag) & (mag) & (mag) & & (\arcsec)\\ 
\hline
51668.89 & -5143 & $\ldots$ & 22.258(0.305) & 22.208(0.274) & 21.924(0.303) & $\ldots$ & SDSS & 1.0\\
56048.57 & -764 & $\ldots$ & 21.635(0.162) & 21.459(0.173) & $\ldots$ & $\ldots$ & NOT+ALFOSC & 0.9\\
56072.64 & -740 & $\ldots$ & 21.637(0.152) & 21.255(0.106) & 21.275(0.123) & $\ldots$ & GTC+OSIRIS & 1.0\\
56147.54 & -665 & $\ldots$ & 21.856(0.211) & 21.562(0.117) & 21.423(0.240) & $\ldots$ & NTT+EFOSC2 & 1.3\\
56385.88 & -426 & $\ldots$ & $\ldots$ & $\ldots$ & $\ldots$ & 21.402(0.050) & CTIO4m+DECam & 1.0\\
56685.73 & -127 & $\ldots$ & 20.973(0.266) & 20.707(0.106) & $\ldots$ & $\ldots$ & NOT+ALFOSC & 1.4\\
56698.72 & -114 & $\ldots$ & 21.253(0.141) & 21.042(0.083) & $\ldots$ & $\ldots$ & NOT+ALFOSC & 1.0\\
56769.68 & -43 & 21.915(0.146) & 20.880(0.064) & 20.625(0.030) & 20.652(0.032) & 20.533(0.072) & CTIO4m+DECam & 1.0\\
56826.39 & 14 & $\ldots$ & 17.128(0.020) & 16.914(0.013) & 17.005(0.019) & $\ldots$ & LT+IO:O & 0.9\\
56828.40 & 16 & $\ldots$ & 17.133(0.017) & 16.901(0.012) & 17.028(0.009) & $\ldots$ & LT+IO:O & 0.9\\
56833.40 & 21 & $\ldots$ & 17.315(0.028) & 16.972(0.011) & 17.125(0.009) & $\ldots$ & LT+IO:O & 1.4\\
56833.41 & 21 & $\ldots$ & 17.293(0.012) & 17.022(0.008) & 17.142(0.010) & $\ldots$ & WHT+ACAM & 1.5\\
56841.41 & 29 & $\ldots$ & 17.813(0.022) & 17.469(0.016) & 17.600(0.015) & $\ldots$ & LT+IO:O & 1.7\\
56851.41 & 39 & $\ldots$ & 18.585(0.033) & 18.091(0.024) & 18.060(0.026) & $\ldots$ & LT+IO:O & 0.6\\
56856.39 & 44 & $\ldots$ & 18.634(0.020) & 18.210(0.013) & 18.150(0.022) & $\ldots$ & LT+IO:O & 0.9\\
56857.42 & 45 & 19.522(0.195) & $\ldots$ & 18.269(0.019) & 18.203(0.020) & $\ldots$ & AS1.82m+AFOSC & 1.8\\
56864.39 & 52 & $\ldots$ & 18.630(0.032) & 18.180(0.019) & 18.127(0.021) & $\ldots$ & LT+IO:O & 0.7\\
56880.41 & 68 & $\ldots$ & 18.538(0.061) & 17.947(0.029) & 17.837(0.023) & $\ldots$ & LT+IO:O & 1.2\\
56881.50 & 69 & $\ldots$ & $\ldots$ & $\ldots$ & $\ldots$ & 17.767(0.021) & CTIO4m+DECam & 1.2\\
56887.38 & 75 & $\ldots$ & 18.231(0.022) & 17.742(0.015) & 17.647(0.013) & $\ldots$ & LT+IO:O & 1.0\\
56895.37 & 83 & $\ldots$ & 18.171(0.022) & 17.662(0.018) & 17.604(0.015) & $\ldots$ & LT+IO:O & 0.9\\
\hline
\end{tabular}
\label{table:phot_gri_hunt248}
\tablefoot{Seeing is derived from the image most similar to \textit{r}-band. Telescopes are detailed in Table~\ref{table:phot_UBVRI_hunt248}.}
\end{table*}
}

\onltab{
\begin{table*}[h]
\caption{Near-IR \textit{JHK} photometry for SNHunt248 with the errors given in brackets.}
\centering
\begin{tabular}{ccccccc}
\hline
\hline
JD & $t$ & $m_{J}$ & $m_{H}$ & $m_{K}$ & Telescope+instrument & Seeing\\
(2400000+) & (d) & (mag) & (mag) & (mag) & & (\arcsec)\\ 
\hline
56053.65 & -759 & 20.555(0.135) & 20.321(0.237) & 20.159(0.264) & WHT+LIRIS & 1.0\\
56818.58 & 6 & 17.129(0.039) & 16.950(0.036) & 16.764(0.057) & NOT+NOTCam & 0.8\\
56848.52 & 36 & 17.338(0.040) & 17.189(0.041) & 16.986(0.053) & NOT+NOTCam & 0.9\\
56876.44 & 64 & 17.177(0.048) & 16.944(0.047) & 16.835(0.095) & NOT+NOTCam & 0.8\\
56905.42 & 93 & 16.630(0.044) & 16.451(0.091) & $\ldots$ & NOT+NOTCam & 1.2\\
57028.77 & 216 & 19.900(0.088) & 19.088(0.056) & 18.083(0.055) & NOT+NOTCam & 0.7\\
57058.76 & 246 & 21.121(0.219) & 19.711(0.086) & 18.669(0.073) & NOT+NOTCam & 0.6\\
57087.70 & 275 & 21.200(0.143) & 20.151(0.151) & 19.121(0.110) & NOT+NOTCam & 0.7\\
57118.58 & 306 & $>$21.3 & 20.418(0.241) & 19.292(0.100) & NOT+NOTCam & 0.9\\
57144.66 & 332 & $\ldots$ & $>$20.8 & 19.640(0.153) & NOT+NOTCam & 0.6\\
\hline
\end{tabular}
\label{table:phot_JHK_hunt248}
\tablefoot{Seeing is derived from the image most similar to \textit{K}-band. Telescopes are detailed in Table~\ref{table:phot_UBVRI_hunt248}.}
\end{table*}
}

\onltab{
\begin{table*}[h]
\caption{Log of spectroscopic observations for SNHunt248.}
\centering
\begin{tabular}{cccccccc}
\hline
\hline
JD & $t$ & Grism/grating & Slit & R & R & Exp. time & Telescope+instrument \\
(2400000+) & (d) & & (\arcsec) & ($\lambda/\Delta\lambda$) & (km s$^{-1}$) & (s) & \\ 
\hline
56817.5 & 5 & \#7 & 1.0 & 650 & 460 & 1800 & NOT+ALFOSC \\
56829.5 & 17 & \#7, \#8 & 1.3 & 500, 770 & 600, 390 & 1800 & NOT+ALFOSC \\
56834.6 & 22 & R300B, R158R & 1.0 & 1100, 970 & 270, 310 & 900 & WHT+ISIS \\
56840.4 & 28 & \#4 & 1.3 & 270 & 1100 & 2400 & NOT+ALFOSC \\
56874.4 & 62 & \#7, \#8 & 1.0 & 650, 1000 & 460, 300 & 2400 & NOT+ALFOSC \\
56882.4 & 70 & LR-B & 1.0 & 590 & 510 & 2700 & TNG+DOLORES \\
56889.3 & 77 & R500B, R500R & 1.0 & 320, 350 & 940, 860 & 1800 & GTC+OSIRIS \\
56894.4 & 82 & \#4 & 1.0 & 360 & 830 & 2400 & NOT+ALFOSC \\
56906.4 & 94 & \#4 & 1.3 & 270 & 1100 & 1800 & NOT+ALFOSC \\
56915.5 & 103 & Red & 1.5 & 26000 & 12 & 2300 & VLT+UVES \\
57068.7 & 256 & R316R & 1.5 & 1300 & 230 & 5400 & WHT+ISIS \\
57130.8 & 318 & 300V & 1.0 & 440 & 680 & 1800 & VLT+FORS2 \\
\hline
\end{tabular}
\label{table:spect_hunt248}
\tablefoot{Telescopes are detailed in Table~\ref{table:phot_UBVRI_hunt248}.}
\end{table*}
}


\begin{thebibliography}{}


\bibitem[de Jager(1998)]{dejager98} de Jager, C.\ 1998, \aapr, 8, 145 
\bibitem[Drake et al.(2009)]{drake09} Drake, A.~J., Djorgovski, S.~G., Mahabal, A., et al.\ 2009, \apj, 696, 870 
\bibitem[Fraser et al.(2013)]{fraser13} Fraser, M., Inserra, C., Jerkstrand, A., et al.\ 2013, \mnras, 433, 1312 
\bibitem[Fraser et al.(2015)]{fraser15} Fraser, M., Kotak, R., Pastorello, A., et al.\ 2015, arXiv:1502.06033 
\bibitem[Humphreys \& Davidson(1994)]{humphreys94} Humphreys, R.~M., \& Davidson, K.\ 1994, \pasp, 106, 1025 
\bibitem[Jester et al.(2005)]{jester05} Jester, S., Schneider, D.~P., Richards, G.~T., et al.\ 2005, \aj, 130, 873 
\bibitem[Kankare et al.(2012)]{kankare12} Kankare, E., Ergon, M., Bufano, F., et al.\ 2012, \mnras, 424, 855 
\bibitem[Lagadec et al.(2011)]{lagadec11} Lagadec, E., Zijlstra, A.~A., Oudmaijer, R.~D., et al.\ 2011, \aap, 534, L10 
\bibitem[Mauerhan et al.(2015)]{mauerhan15} Mauerhan, J.~C., Van Dyk, S.~D., Graham, M.~L., et al.\ 2015, \mnras, 447, 1922 
\bibitem[Pastorello et al.(2013)]{pastorello13} Pastorello, A., Cappellaro, E., Inserra, C., et al.\ 2013, \apj, 767, 1 
\bibitem[Schlafly \& Finkbeiner(2011)]{schlafly11} Schlafly, E.~F., \& Finkbeiner, D.~P.\ 2011, \apj, 737, 103 
\bibitem[Smith et al.(2011)]{smith11} Smith, N., Li, W., Silverman, J.~M., Ganeshalingam, M., \& Filippenko, A.~V.\ 2011, \mnras, 415, 773 
\bibitem[Tully et al.(2009)]{tully09} Tully, R.~B., Rizzi, L., Shaya, E.~J., et al.\ 2009, \aj, 138, 323 
\bibitem[Valenti et al.(2011)]{valenti11} Valenti, S., Fraser, M., Benetti, S., et al.\ 2011, \mnras, 416, 3138 
\bibitem[Van Dyk et al.(2000)]{vandyk00} Van Dyk, S.~D., Peng, C.~Y., King, J.~Y., et al.\ 2000, \pasp, 112, 1532 
\bibitem[Zheng et al.(2014)]{zheng14} Zheng, W., Filippenko, A.~V., Graham, M.~L., Mauerhan, J., \& Van Dyk, S.~D.\ 2014, The Astronomer's Telegram, 6206, 1 


\end{thebibliography}
\end{document}